\documentclass[12pt,fleqn]{article}
\usepackage{amsmath}

\begin{document}

\title{\bf The short pulse equation\\
is integrable}

\author{{\sc Anton Sakovich$^{1)}$, Sergei
Sakovich$^{2)}$}\\[16pt] \small{$^{1)}$Department of Physics,
Belarusian State University,}\\[-6pt] \small{220080 Minsk,
Belarus. E-mail: ant.s@tut.by}\\[6pt] \small{$^{2)}$Institute of
Physics, National Academy of Sciences,}\\[-6pt] \small{220072
Minsk, Belarus. E-mail: saks@tut.by}}

\date{}

\maketitle

\begin{abstract}
We prove that the Sch\"{a}fer--Wayne short pulse equation (SPE)
which describes the propagation of ultra-short optical pulses in
nonlinear media is integrable. First, we discover a Lax pair of
the SPE which turns out to be of the Wadati--Konno--Ichikawa type.
Second, we construct a chain of transformations which relates the
SPE with the sine-Gordon equation.
\end{abstract}

\medskip

\section{Introduction}

Recently, Sch\"{a}fer and Wayne \cite{SW} derived the short pulse
equation (SPE) as a model equation, alternative to the nonlinear
Schr\"{o}dinger equation (NLSE), to approximate the evolution of
very short optical pulses in nonlinear media. More recently,
Chung, Jones, Sch\"{a}fer and Wayne \cite{CJSW} proved numerically
that, as the pulse length shortens, the NLSE approximation becomes
less and less accurate while the SPE provides a better and better
approximation to the solution of Maxwell's equation.

In the present paper, we study this interesting new nonlinear
equation, the SPE, from the standpoint of its integrability. The
NLSE is well known as one of the numerous nonlinear equations
integrable by the inverse scattering transform technique
\cite{AS}. Therefore it is natural to ask whether the SPE, as an
ultra-short pulse alternative to the NLSE, is integrable as well,
or one has to rely mainly on numerical techniques when studying
the SPE. We prove that the SPE is integrable. In Section~\ref{s2},
we find that the SPE possesses a Lax pair of the
Wadati--Konno--Ichikawa (WKI) type \cite{WKI}. Then, in
Section~\ref{s3}, we show how the SPE can be transformed into the
well-known integrable sine-Gordon equation \cite{AS}.
Section~\ref{s4} summarizes the results.

We study the SPE in the form
\begin{equation} \label{e1}
u_{xt} = u + \tfrac{1}{6} \left( u^3 \right)_{xx}
\end{equation}
which is related to the original form used in \cite{SW,CJSW} by
scale transformations of the dependent and independent variables.

\section{Zero-curvature representation} \label{s2}

Let us try to find a zero-curvature representation (ZCR) of the
SPE \eqref{e1},
\begin{equation} \label{e2}
D_t X - D_x T + [ X , T ] = 0 ,
\end{equation}
which is the compatibility condition of the overdetermined linear
system
\begin{align}
\Psi_x & = X \Psi , \label{e3} \\ \Psi_t & = T \Psi , \label{e4}
\end{align}
where $X$ and $T$ are $n \times n$ matrix functions of $u$ and its
derivatives, $\Psi (x,t)$ is an $n$-component column, $D_x$ and
$D_t$ denote the total derivatives, and the square brackets denote
the matrix commutator.

Taking for simplicity
\begin{equation} \label{e5}
X = u_x A + B , \qquad T = T ( u , u_x ) ,
\end{equation}
with constant matrices $A$ and $B$, and replacing $u_{xt}$ by the
right-hand side of \eqref{e1}, we find from \eqref{e2} that
\begin{equation} \label{e6}
T = \tfrac{1}{2} u^2 u_x A + S ( u ) ,
\end{equation}
where the matrix function $S(u)$ satisfies the equations
\begin{equation} \label{e7}
d S / d u = \left[ A , S - \tfrac{1}{2} u^2 B \right] , \qquad u A
+ [ B , S ] = 0 .
\end{equation}

Setting $S(u)$ to be the simplest nontrivial polynomial,
\begin{equation} \label{e8}
S = \tfrac{1}{2} u^2 B + u P + Q
\end{equation}
with constant matrices $P$ and $Q$, we have from \eqref{e7}
\begin{equation} \label{e9}
A = - [ B , P ]
\end{equation}
and
\begin{equation} \label{e10}
\bigl[ P , [ B , P ] \bigr] = B , \qquad [ B , Q ] = 0 , \qquad
\bigl[ Q , [ B , P ] \bigr] = P .
\end{equation}

A nontrivial solution of the commutator equations \eqref{e10} can
be easily found in traceless $2 \times 2$ matrices $B$, $P$ and
$Q$. Using this solution, together with \eqref{e5}, \eqref{e6},
\eqref{e8} and \eqref{e9}, and simplifying the expressions for $X$
and $T$ by $X \mapsto G X G^{-1}$ and $T \mapsto G T G^{-1}$ with
a constant matrix $G$, we obtain the following:
\begin{equation} \label{e11}
X =
\begin{pmatrix}
\lambda & \lambda u_x \\ \lambda u_x & - \lambda
\end{pmatrix}
\end{equation}
and
\begin{equation} \label{e12}
T =
\begin{pmatrix}
\frac{\lambda}{2} u^2 + \frac{1}{4 \lambda} & \frac{\lambda}{2}
u^2 u_x - \frac{1}{2} u \\[4pt] \frac{\lambda}{2} u^2 u_x +
\frac{1}{2} u & - \frac{\lambda}{2} u^2 - \frac{1}{4 \lambda}
\end{pmatrix}
,
\end{equation}
where $\lambda$ is an arbitrary nonzero constant.

Consequently, the SPE \eqref{e1} possesses the ZCR \eqref{e2}, or
the Lax pair \eqref{e3} and \eqref{e4}, with the matrices $X$ and
$T$ determined by \eqref{e11} and \eqref{e12}. The linear problem
\eqref{e3} with the matrix $X$ \eqref{e11} is a spectral problem
of the WKI type \cite{WKI} with respect to $u_x$. Note, however,
that the matrix $T$ \eqref{e12} is nonlocal with respect to $u_x$
and contains $\lambda^{-1}$. The WKI-type ZCRs with matrices $T$
containing negative powers of the spectral parameter were studied
in \cite{BPT}. Nevertheless, the nonlinear equation we study, the
SPE \eqref{e1}, did not appear in the literature prior to
\cite{SW,CJSW}, as far as we know.

We have already pointed out that the form \eqref{e1} of the SPE is
related to the original form used in \cite{SW,CJSW} by scale
transformations of variables. Note, however, that some
complex-valued scalings of variables are required in order to
change the sign at the nonlinear term in \eqref{e1}. Therefore, if
one does not accept complex-valued transformations (say, as
nonphysical) but is interested in the form
\begin{equation} \label{e13}
u_{xt} = u - \tfrac{1}{6} \left( u^3 \right)_{xx}
\end{equation}
of the SPE, one needs to use the ZCR \eqref{e2} with the matrices
\begin{equation} \label{e14}
\begin{split}
X_{(-)} & =
\begin{pmatrix}
\lambda & \lambda u_x \\ - \lambda u_x & - \lambda
\end{pmatrix}
, \\[4pt] T_{(-)} & =
\begin{pmatrix}
- \frac{\lambda}{2} u^2 + \frac{1}{4 \lambda} & -
\frac{\lambda}{2} u^2 u_x - \frac{1}{2} u \\[4pt]
\frac{\lambda}{2} u^2 u_x - \frac{1}{2} u & \frac{\lambda}{2} u^2
- \frac{1}{4 \lambda}
\end{pmatrix}
\end{split}
\end{equation}
for the `minus' form \eqref{e13} of the SPE. All this is
completely analogous to the situation with the `plus' and `minus'
forms of the NLSE \cite{AS}.

\section{Equivalence transformation} \label{s3}

Quite often an interesting newly-found nonlinear equation turns
out to be equivalent to a well-studied old one through some chain
of transformations. Let us see that the SPE \eqref{e1} is not an
exception in this respect. Of course, since the SPE possesses a
WKI-type Lax pair, it is possible to use the known interrelation
\cite{I,WS} between spectral problems of the WKI type and the
Ablowitz--Kaup--Newell--Segur (AKNS) type, in order to find which
equation possessing an AKNS-type Lax pair corresponds to the SPE.
However, we follow a different way when trying to transform the
SPE into some old equation: we use a generalized symmetry of the
SPE as an indicator of required transformations. This heuristic
method is based on the facts that any generalized symmetry is
formally represented by an evolution equation \cite{O} and that
the transformations of evolution equations are studied much better
than transformations of other types of equations.

It is easy to find directly that the SPE \eqref{e1} admits a
third-order generalized symmetry, which is written in the
evolutionary form as
\begin{equation} \label{e15}
u_s = \left( u_x \left( u_x^2 + 1 \right)^{-1/2} \right)_{xx} ,
\end{equation}
where the subscript $s$ denotes the derivative with respect to the
group parameter $s$. This symmetry can also be obtained using the
recursion operator
\begin{equation} \label{e16}
R = D_x^2 \cdot u_{xx}^{-1} D_x \cdot \left( u_x^2 + 1
\right)^{1/2} D_x^{-1} \cdot \left( u_x^2 + 1 \right)^{-3/2}
u_{xx}
\end{equation}
easily derivable from the matrix $X$ \eqref{e11} by the cyclic
basis technique \cite{S1,S2}. This operator generates the symmetry
\eqref{e15} from the trivial symmetry $u_s = 0$, not from the
translational symmetry $u_s = u_x$, because $R u_x = 0$. In what
follows, however, we do not study the recursion operator
\eqref{e16}.

We consider the generalized symmetry \eqref{e15} as a third-order
evolution equation in $u(x,t,s)$, and try to transform it into a
simple known evolution equation, following the way used in
\cite{S2,S3}.

First of all, we find the separant $\partial f / \partial u_{xxx}$
of the equation \eqref{e15}, where $f$ denotes the equation's
right-hand side, introduce the new dependent variable $v(x,t,s) =
\left( \partial f / \partial u_{xxx} \right)^{1/3}$, and thus
obtain the transformation
\begin{equation} \label{e17}
v ( x , t , s ) = \left( u_x^2 + 1 \right)^{-1/2}
\end{equation}
which relates the equation \eqref{e15} with the equation
\begin{equation} \label{e18}
v_s = v^3 v_{xxx} + \frac{3 v^2 v_x v_{xx}}{1 - v^2} + \frac{3 v^3
v_x^3}{\left( 1 - v^2 \right)^2}
\end{equation}
possessing the separant $v^3$.

Then, we introduce the new independent and dependent variables,
$y$ and $w(y,t,s)$, such that
\begin{equation} \label{e19}
x = w ( y , t , s ) , \qquad v ( x , t , s ) = w_y ( y , t , s ) .
\end{equation}
This transformation \eqref{e19} relates the equation \eqref{e18}
with the constant-separant equation
\begin{equation} \label{e20}
w_s = w_{yyy} + \frac{3 w_y w_{yy}^2}{2 \left( 1 - w_y^2 \right)}
+ \phi ( t , s ) w_y ,
\end{equation}
where the arbitrary function $\phi(t,s)$ appeared as a `constant'
of the integration over $y$. This inessential function $\phi(t,s)$
can be turned into zero by the redefinition of $y$, $y \mapsto y +
\int \phi(t,s) \: d s$, which does not change $v(x,t,s)$.

At last, trying to eliminate the second-order derivative terms
from the transformed equation, we find that the new dependent
variable
\begin{equation} \label{e21}
z ( y , t , s ) = \arccos ( w_y )
\end{equation}
transforms the equation \eqref{e20} (where we have already set
$\phi(t,s)=0$) into the potential modified Korteweg--de~Vries
equation
\begin{equation} \label{e22}
z_s = z_{yyy} + \tfrac{1}{2} z_y^3 .
\end{equation}

We have transformed the generalized symmetry \eqref{e15} of the
SPE into the simple known evolution equation \eqref{e22}. The
equation \eqref{e22} represents a generalized symmetry of the
sine-Gordon equation $z_{yt} = \sin z$ \cite{O}, but not only of
it: for example, the equations $z_{yt} = \cos z$, $z_{yt} = \exp
(\pm \mathrm{i} z)$ and $z_{yt} = 0$ also admit the generalized
symmetry \eqref{e22}. In order to see which equation possessing
the generalized symmetry \eqref{e22} corresponds to the SPE, we
apply the chain of transformations \eqref{e17}, \eqref{e19} and
\eqref{e21} directly to the SPE \eqref{e1}, and actually obtain
the sine-Gordon equation $z_{yt} = \sin z$. Note that, when making
the transformation \eqref{e19}, we neglect the `constant' of the
integration over $y$, like we did this in the equation \eqref{e20}
and for the same reason.

Consequently, the SPE \eqref{e1} and the sine-Gordon equation
\begin{equation} \label{e23}
z_{yt} = \sin z
\end{equation}
are equivalent to each other through the chain of transformations
\begin{gather}
v ( x , t ) = \left( u_x^2 + 1 \right)^{-1/2} ; \label{e24} \\ x =
w ( y , t ) , \qquad v ( x , t ) = w_y ( y , t ) ; \label{e25} \\
z ( y , t ) = \arccos ( w_y ) . \label{e26}
\end{gather}

\section{Conclusion} \label{s4}

In this paper, we studied the integrability of the short pulse
equation (SPE) which describes the propagation of ultra-short
optical pulses in nonlinear media. We found a Lax pair of the SPE,
of the Wadati--Konno--Ichikawa type. This makes possible to use
the powerful inverse scattering transform technique for obtaining
and analyzing solutions of the SPE. Also we found how to transform
the SPE into the well-studied sine-Gordon equation. This makes
possible to derive solutions and properties of the SPE from known
solutions and properties of the sine-Gordon equation. We believe
that these two possibilities to study the SPE analytically,
discovered in this paper, will be valuable for soliton theory and
nonlinear optics.

\end{document}